\documentclass[aip,reprint]{revtex4-1}            

\usepackage{amsmath}
\usepackage{graphicx} 
\usepackage{chemmacros}
\newcommand{\e}{\mathrm{e}}     
\newcommand{\dd}{\mathrm{d}}  
\newcommand{\DD}{\mathcal{D}} 
\newcommand{\ii}{\mathrm i}    

\newcommand{\rr}{{\bf r }}   
\newcommand{\kk}{{\bf k }}  
\newcommand{\RR}{{\bf R }}   

\newcommand{\lB}{l_{\rm B}}

\newcommand{\np}{n_{\rm p}} 
\newcommand{\nT}{n_{\rm T}} 
\newcommand{\nbound}{n_{\rm AB} }
\newcommand{\nfree}{n_{\rm A} }

\newcommand{\rhob}{\rho_{\rm b}}
\newcommand{\rhoT}{\rho_{\rm T} }

\newcommand{\lambdaT}{\lambda_{\rm th} }

\newcommand{\diag}{{\rm diag}}

\usepackage{color}

\makeatletter
\def\@email#1#2{%
 \endgroup
 \patchcmd{\titleblock@produce}
  {\frontmatter@RRAPformat}
  {\frontmatter@RRAPformat{\produce@RRAP{*#1\href{mailto:#2}{#2}}}\frontmatter@RRAPformat}
  {}{}
}%

\begin{document}

\title{Field theory description of ion association in re-entrant phase separation of polyampholytes}

\author{Jonas~Wess\'en}
\affiliation{Department of Biochemistry, University of Toronto, 1 King’s College Circle, Toronto, Ontario M5S 1A8, Canada}

\author{Tanmoy~Pal}

\affiliation{Department of Biochemistry, University of Toronto, 1 King’s College Circle, Toronto, Ontario M5S 1A8, Canada}

\author{Hue~Sun~Chan}

\affiliation{Department of Biochemistry, University of Toronto, 1 King’s College Circle, Toronto, Ontario M5S 1A8, Canada}

\email{tanmoy.pal@utoronto.ca}
\email{jonas.wessen@utoronto.ca}
\date{\today}

\begin{abstract}
Phase separation of several different overall neutral 
polyampholyte species (with zero net charge) is studied in solution 
with two oppositely charged ion species that can form ion-pairs through 
an association reaction. A field theory description of the system, that 
treats polyampholyte charge sequence dependent electrostatic 
interactions as well as excluded volume effects, is hereby given. 
Interestingly, analysis of the model using random phase approximation 
and field theoretic simulation consistently show evidence of a 
re-entrant polyampholyte phase separation at high ion concentrations 
when there is an overall decrease of volume upon ion-association. As an
illustration of the ramifications of our theoretical framework, several 
polyampholyte concentration vs ion concentration phase diagrams under
constant temperature conditions are presented to elucidate the dependence 
of phase separation behavior on polyampholyte sequence charge pattern as
well as ion-pair dissociation constant, volumetric effects on ion association, 
solvent quality, and temperature.
\end{abstract}


\maketitle 

\section{Introduction}
Numerous studies in the past decade have indicated that liquid-liquid phase
separation (LLPS) of intrinsically disordered proteins (IDPs) or intrinsically
disordered regions (IDRs) of proteins, often in the presence of nucleic acids
and folded protein domains,
is a critical physical mechanism behind the formation of biologically functional
membraneless organelles such as the nucleolus, cajal bodies and stress
granules\cite{BrangwynneHyman2009, Rosen12, McKnight12, CellBiol, Nott2015,
BananiLeeHymanRosen2017, ALBERTI2017R1097, LiChavaliPancsaChavaliBabu,
MolliexTemirovLeeCoughlinKanagarajKimMittagTaylor, NatPhys,
Chong2016, Banani2017, mingjie2020}. 
Electrostatics, among other multivalent interactions, is one of the major 
drivers of the
biological LLPS because IDP/IDRs generally contain charged amino acid residues
in their composition and are often polyampholytic in nature \cite{uversky2002,
cosb15, Robert-Julie, SumanPNAS}. In addition to a polyampholyte sequence's
charge pattern---or more generally its charge composition---and 
concentration \cite{SumanJPCB2018, JonasJPCB2021, LinPRL,
McCartyDelaneyDanielsenFredricksonShea2019}, the phase separation propensity of
a specific polyampholyte depends on the condition of the solution determined by
such factors as temperature \cite{rolandJACS2019, jeetainACS}, pressure
\cite{rolandJACS2019, roland20,FetahajJACP2021}, pH \cite{WeberJulicher2020},
as well as the concentration and type of salts present \cite{FetahajJACP2021,
EspinosaJCP2021}. Because of the charged nature of the polyampholytes,
electrostatic interactions from the ion pairs in the solution affects its LLPS
\cite{LinPRL, Kraineretal2020, EspinosaJCP2021,FetahajJACP2021} and could be a
controlling factor in its bio-engineering \cite{Hong2020}. In general, an
oppositely charged pair of ions can stay as solvated ions or they can form a
chemically distinct complex, e.g.~solvent-shared ion-pair, contact ion-pair,
through association reactions depending on the solution condition. Although the
effects of ion-strength on polyampholyte LLPS have been addressed by several
analytical and computational studies, the consequences of ion-association has
largely been unexplored.

In general, chemical reactions in biomolecular condensates are of broad
interest because the regulation of biochemical reactions is one of
several major biologically relevant functions of biomolecular
condensates\cite{AlbertiGladfelterMittag2019}. Indeed, several recent 
computational studies have addressed phase separation in chemically 
reactive environments. For instance,
the pH dependence of LLPS was studied by Adame-Arana \textit{et al}.~in a
set-up where the net charge on the phase separating macromolecules is
chemically coupled to the self-ionization of water \cite{WeberJulicher2020}.
An investigation by Lin \textit{et al}.~elucidated the role of complex formation
between the SynGAP and PSD-95 molecules in the LLPS of their mixture
\cite{LinWuJiaZhangChan2021}. The study by Bartolucci \textit{et
al}.~considered both equilibrium and fuel-driven phase separation of a
polymeric component undergoing an internal molecular
transition\cite{Bartolucci2021}. These studies provided valuable insights.
However, they treated relevant interactions only up to mean-field theory 
(MFT) and did not incorporate amino acid sequence of the polyampholytes 
explicitly. Sequence specificity, and generally phase-separation driven 
by electrostatic interactions, are inaccessible to the aforementioned
MFT approaches because of the non-neglible contributions from
fluctuations\cite{PopovLeeFredrickson2007,LinPRL}; but these effects
are physically and biochemically important. One of the goals of our
present work is to develop theoretical approaches that allow these effects
to be tackled.

Pinpointing the exact roles played by all the physical interactions affecting
\textit{in vivo} LLPS, or even its simplified \textit{in vitro} counterpart,
could be immensely difficult. For analytical and computational tractability, we
consider a simple model where a polyampholyte species is phase-separating in
the presence of two oppositely charged chemically reacting ion species ${\rm
A}^+$ and ${\rm B}^-$. We assume that the concentrations of ${\rm A}^+$ and
${\rm B}^-$ are in thermal equilibrium with the concentration of their
charge-neutral product ${\rm AB}$, following the balance equation
\begin{equation}
\label{eq:balance_equation}
\ch{ A^+ + B^- <=>[$K\sb{\rm a}$][$K\sb{\rm d}$] AB }.
\end{equation}
The dissociation\slash association constants $K_{\rm d}$/$K_{\rm a}$ 
in Eq.~\eqref{eq:balance_equation} are defined by
\begin{equation} \label{eq:Kd_definition}
K_{\rm d} = \frac{1}{K_{\rm a} } = \frac{[ { \rm A }^+] [{\rm B}^- ] }{ [{ \rm AB}] },
\end{equation}
where $[X]$ is the equilibrium concentration of the species $X$ ($= {\rm A}^+,
{\rm B}^-, {\rm AB}$). The reaction \eqref{eq:balance_equation} could be used
to describe several chemical processes including self-ionization of water,
dissociation of weak organic acids (e.g.~formic acid, acetic acid or carbonic
acid) in the solution, ion association in concentrated solutions of
electrolytes at physical temperatures, at non-polarity solvents or at low
temperatures \cite{Valeriani2010}. The dissociation constant $K_{\rm d}$ of a
chemical reaction is an experimentally measurable observable whose value can
indicate the chemical state of the solution. If the initial reactant
concentrations are lower than $K_{\rm d}$, most of the ion pairs are expected
to be in the dissolved state which will result in screening of the
polyampholyte's electrostatic interactions. On the other hand, when the
reactant concentrations are considerably higher than $K_{\rm d}$, most of the
ion pairs are expected to be in the charge-neutral complex state which will
affect the configuration entropy of the polyampholyte by modulating the
effective excluded volume through steric repulsion. In addition, ion-pair
association is often accompanied by a change in volume
\cite{SpiroReveszLee1968, WarkHsiaSon2008, Marcus1983}. Any such volume change
might have an important effect---in addition to the electrostatic
screening effects of the ions---on polyampholyte conformation at high reactant
concentrations. Thus, the chemical state of the solution determined by
Eq.~\eqref{eq:balance_equation} has potential to dictate LLPS behavior of the
polyampholyte species. 

At high salt concentrations, non-electrostatic interactions are expected to
play a major role in determining LLPS behavior. Indeed, in a recent explicit
chain molecular dynamics (MD) study, 
phase separation behaviour of several proteins at high salt concentrations were
attributed to non-electrostatic interactions such as hydrophobic
interactions\cite{Kraineretal2020}. In a different study, the importance of the 
excluded volume interaction from PEG crowding agents was
highlighted in a system with a phase separating protein that lack hydrophobic
amino acids in its sequence \cite{ParkFredrickson2020}. A model that includes
both ion-association along with any volume change upon ion association and
explicit residue level electrostatics of the phase separating polyampholyte
thus offers a unique possibility of capturing many of the diverse results
mentioned above in an unified set-up, at least qualitatively. 

With that in mind, here we adopt a trade-off between complexity and
analytical/computational tractability by introducing a simple field theory model
where molecular species in the model interact via excluded volume and 
explicit sequence dependent electrostatic interactions. 
Specifically, we introduce a bare dissociation constant (corresponding
to the dissociation constant in a solution consisting only of A$^+$, B$^-$, 
and AB)
as a control parameter for the non-electrostatic energy gain associated with
ion-pairing. To account for the volume change upon association, we introduce
further a relative excluded volume factor $\gamma$ of the product ${\rm AB}$
in Eq.~\eqref{eq:balance_equation} with respect to the reactants ${\rm A}$ and
${\rm B}$. We study the model using MFT, random phase approximation 
(RPA)---where
relevant Gaussian level fluctuations are included above the MFT\cite{Gonzales-Mozuelos1994}, and fully
fluctuating field theoretic simulation (FTS). We expect the model 
to be useful as a base for studying specific systems with suitable
modifications. 

The structure of this article is as follows. In Sec.~\ref{sec:model_def}, we
introduce our model and derive its corresponding field theory representation in
Sec.~\ref{sec:deriving_field_theory}. The model is studied analytically in
Sec.~\ref{sec:analytical} using MFT and RPA, and then using FTS in
Sec.~\ref{sec:FTS}. Numerical results obtained from the approximate analytical
calculations and from FTS are shown and compared in Sec.~\ref{sec:results}, and
concluding remarks are given in Sec.~\ref{sec:conclusions}. 

\section{Model definition} \label{sec:model_def}

\begin{figure}
\includegraphics{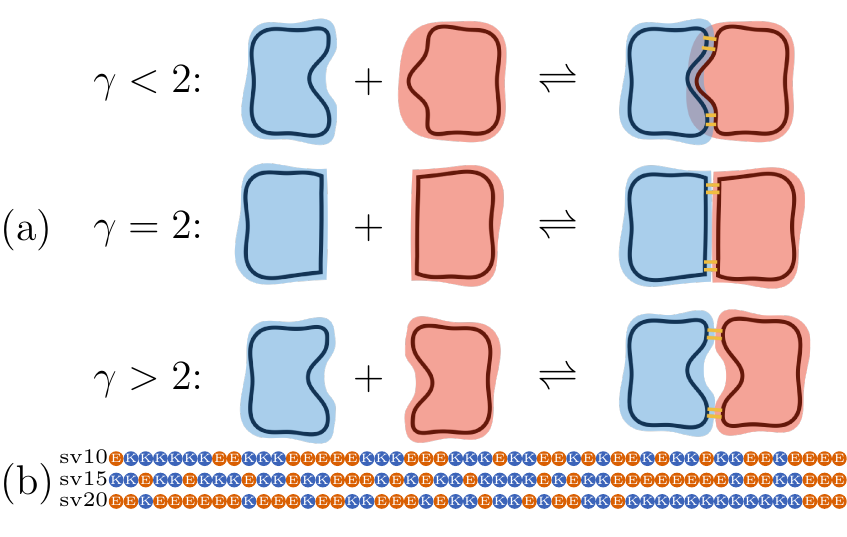}
\caption{(a) Schematic illustration of the effective volume change 
upon ion-association, modeled by the parameter $\gamma$. The shading represents volumetric effect beyond that of the
van der Waals volume represented by the solid outlines of the solutes.
(b) Charge sequences considered in this work. Blue (`K') and red (`E') 
beads correspond to charges $+1$ and $-1$, respectively.
} 
\label{fig:cartoons}
\end{figure}

Our system of consideration contains $\np$ linear polymers of identical composition, each consisting of $N$ residues with electric charges $\sigma_{\alpha}$, $\alpha=1,\dots,N$. Effects from solvent molecules are assumed to be implicitly encoded in the microscopic interaction parameters. All electric charges in this work are given in units of the elementary protonic charge, and we restrict our calculations to polymers with zero net electric charge, $\sum_{\alpha=1}^N \sigma_{\alpha} = 0$.  The `K'--`E' sequences (Fig.~\ref{fig:cartoons} (b)) considered here are representative of the ``sv sequences''\cite{DasPappu2013PNAS} used extensively to study LLPS in computational models\cite{DasAminLinChan2018, PRE2021Pal,Lin2021,McCartyDelaneyDanielsenFredricksonShea2019,SumanJPCB2018,JonasJPCB2021}. An analytically derived single chain property, the sequence charge decoration (SCD) parameter (defined as $\mathrm{SCD} = \sum_{\beta=2}^{N} \sum_{\alpha=1}^{\beta-1} \sigma_{\beta} \sigma_{\alpha} \sqrt{ \left( \beta -\alpha \right)} / N$), could be used to discriminate between otherwise charge neutral `sv' sequences\cite{SawleGosh2015}. The phase separation propensity trends of the ``sv sequences'' is known to correlate well with their SCD parameter values\cite{DasAminLinChan2018,McCartyDelaneyDanielsenFredricksonShea2019,JonasJPCB2021}. A `K'--`E' sequence ($(\mathrm{K}_8 \mathrm{E}_8)_3$) with SCD value -4.290 has recently been seen to be phase separating in an \textit{in vitro} experiment\cite{MadinyaPerry2020}.
Compared to that the SCD values of the sequences studied here are -2.098, -4.349 and -7.374, respectively for sv10, sv15 and sv20.  However, we emphasize that the model presented here is not intended to be quantitatively accurate given that we ignore some effects that are present in experimental systems, e.g.~the
possibility of ion condensation onto the polymer residues due to attractive interactions beyond the simple Coulomb forces\cite{Salehi2016,Friedowitz2018,Ghasemi2020,MadinyaPerry2020,Sing2020,Knoerdel2021}. Consequently, while the trend predicted by our theory is expected to hold for corresponding experimental results, there are uncertainties \cite{NoidReview,Lin2021} in mapping the theoretical variables such as temperature to their experimental counterparts for individual systems.

The system further contains equal amounts of unit positive and negative model ions (denoted ${\rm A}^+$ and ${\rm B}^-$, respectively) that can undergo a pair-wise chemical reaction forming a neutral complex ${\rm AB}$ according to Eq.~\eqref{eq:balance_equation}. The number of ${\rm A}$\slash${\rm B}$ pairs in the dissociated and bound state, $\nfree$ ($=n_{\rm B}$) and $\nbound$ respectively, are thus constrained according to
\begin{equation} \label{eq:nFree_constraint}
\nfree + \nbound = \nT , 
\end{equation}
where $\nT$ denotes the total number of ${\rm A}$ and ${\rm B}$ units in the system.

The position of bead $\alpha$ on polymer $i=1,\dots,\np$ is denoted $\RR_{i,\alpha}$, while the positions of the ${\rm A}^+$-, ${\rm B}^-$- and ${\rm AB}$ particles are represented by $\lbrace \rr_{{\rm A}, i} \rbrace_{i=1}^{\nfree}$, $\lbrace \rr_{{\rm B}, i} \rbrace_{i=1}^{\nfree}$ and $\lbrace \rr_{{\rm AB}, i} \rbrace_{i=1}^{\nbound}$, respectively. Quantities that depend explicitly on molecule positions are indicated by a hat (e.g.~$\hat{H}$). All energies in this work are given in units of the thermal energy $k_{\rm B} T$. The total Hamiltonian is $ \hat{H} = \hat{H}_0 + \hat{H}_1 + \hat{H}_2 + \hat{H}_3$, where
\begin{equation}
\label{eq:interactions}
\begin{aligned}
\hat{H}_0 &= \frac{3}{2 b^2} \sum_{i=1}^{\np} \sum_{\alpha=1}^{N-1} \left( \RR_{i,\alpha+1} - \RR_{i,\alpha} \right)^2 , \\
\hat{H}_1 &= \frac{v}{2} \int \dd \rr \hat{\rho}(\rr)^2 , \\
\hat{H}_2 &= \frac{\lB}{2} \int \dd \rr \int \dd \rr'  \frac{\hat{c}(\rr) \hat{c}(\rr')}{|\rr - \rr'|} . 
\end{aligned}
\end{equation}
These terms provide, respectively, the energies associated with
harmonic chain connectivity, excluded volume repulsion, 
and electrostatic interactions. Here $b$ is the polymer segment length, $v$ is the excluded volume parameter and $\lB = e^2 / 4 \pi \epsilon k_{\rm B} T$ is the Bjerrum length ($v$ and $\lB$ modulate the strengths of the excluded volume- and electrostatic interactions, respectively). We use the segment length $b$ as our unit of length, and thus values of all 
dimensionfull quantities (those that carry a physical unit) 
are given in powers of $b$. We have expressed $\hat{H}_{1,2}$ in terms of the microscopic particle densities and charge densities, defined respectively as
\begin{align} 
\hat{\rho}(\rr) &= \hat{\rho}_{\rm b}(\rr) + \hat{\rho}_{\rm A}(\rr) + \hat{\rho}_{\rm B}(\rr) + \gamma \hat{\rho}_{\rm AB}(\rr) \label{eq:micro_density} , \\
\hat{c}(\rr) &= \hat{c}_{\rm b}(\rr) +  \hat{\rho}_{\rm A}(\rr) - \hat{\rho}_{\rm B}(\rr) ,
\end{align}
where $\hat{\rho}_{\rm b}(\rr)= \sum_{i,\alpha} \Gamma(\rr - \RR_{i,\alpha})$, $\hat{c}_{\rm b}(\rr)= \sum_{i,\alpha} \sigma_{\alpha} \Gamma(\rr - \RR_{i,\alpha})$, $\hat{\rho}_{j}(\rr)= \sum_{i} \Gamma(\rr - \rr_{j,i})$ for $j={\rm A}, {\rm B}, {\rm AB}$ and $\gamma$ represents the relative excluded volume strength of the ${\rm AB}$ complex to a free ion (either ${\rm A}^+$ or ${\rm B}^-$). Individual polymer beads and ions are treated as Gaussian distributions $\Gamma(\rr) = \e^{- \rr^2 / 2 a^2} / (2 \pi a^2)^{3/2}$ with smearing length $a=b/\sqrt{6}$ to regulate ultraviolet divergences arising from contact excluded volume- and electrostatic contact interactions\cite{Wang2010,RigglemanRajeevFredrickson2012}. This Gaussian smearing procedure has also been shown to remedy unphysical binding behaviour in models of highly concentrated polyelectrolyte solutions that combine equilibrium constants with Debye-H{\"u}ckel treatment of electrostatic interactions\cite{Friedowitz2018}.
In this work, we view the Gaussian smearing as a part of the model definition, rather than as an approximation. While leaving the long-range behavior of inter-particle forces unaffected, the Gaussian smearing results in short-range interactions that are considerably ``softer'' than typical hard-sphere or Lennard-Jones particle representations\cite{PRE2021Pal}. Consequently, the model is not expected to capture microscopic phenomena such as strongly oscillating radial distribution functions on short distance scales that typically characterizes liquid phases\cite{Danielsen2019}.

The parameter $\gamma$ in Eq.~\eqref{eq:micro_density} may be interpreted through the effective volume of the ${\rm AB}$ state relative to its dissociated counterpart, with the sign of $\gamma - 2$ corresponding to the sign of ${\rm vol}({\rm AB}) - {\rm vol}({\rm A}^+) - {\rm vol}({\rm B}^-)$ as illustrated schematically in Fig.~\ref{fig:cartoons} (a). The resulting ${\rm AB}$--${\rm AB}$ and ${\rm AB}$--[other monomer $={\rm A}^+$, ${\rm B}^-$, (polymer bead)] excluded volume interaction strengths are scaled as $\gamma^2$ and $\gamma$, respectively, relative to the [other monomer]--[other monomer] excluded volume repulsion strength $v$. At the MFT level of an analogous model that takes the ${\rm AB}$ bound state as an independent system component, these excluded volume interactions result in effective $\chi$-parameters that are scaled by the same powers of $\gamma$, and our framework is therefore related to other approaches in the literature\cite{Cruz1995,Kudlay_macromolecules_2004,Kudlay_JCP_2004} where effective ion size effects are modeled through appropriate $\chi$-parameter variations. Implications of effective ion sizes on counter-ion condensation have also been considered within the context of polyelectrolyte complex coacervation\cite{Ghasemi2020}.

The final piece of the Hamiltonian, $\hat{H}_3$, models the chemical binding of the ${\rm AB}$ complex,
\begin{equation}
\hat{H}_3 = (\nT - \nbound) \epsilon_{\rm bound} ,
\end{equation}
where $-\epsilon_{\rm bound}$ is the decrement in free energy associated with a single ${\rm AB}$ binding, in addition to electrostatic and excluded volume contributions coming from $\hat{H}_{1,2}$. Note that $\epsilon_{\rm bound}$ may include both an energetic component, as well as an entropic component related e.g.~to changes in the number of configuration states, or the entropy release from reduced electrostriction of nearby solvent molecules\cite{Laidler1987}. In the remainder of this article, we find it convenient to trade $\epsilon_{\rm bound}$ in favor of another parameter $K_0$, defined as
\begin{equation}
K_0 = \frac{ \e^{- \epsilon_{\rm bound}} }{ \lambdaT^3 }, 
\end{equation}
where $\lambdaT$ is the thermal de Broglie wavelength. We show in Sec.~\ref{sec:MFT} that $K_0$ corresponds to the mean-field dissociation constant in the dilute limit. This simple physical interpretation of $K_0$ breaks down when 
particle and charge fluctuations are taken into
consideration, as shown in Sec.~\ref{sec:RPA}. 

The partition function of the system,
\begin{equation} \label{eq:partition_function_particle_picture}
Z = \sum_{\nbound = 0}^{\nT} \frac{ \int \lbrace \dd \RR \rbrace \lbrace \dd \rr_{\rm A}\rbrace \lbrace \dd \rr_{\rm B}\rbrace \lbrace \dd \rr_{\rm AB }\rbrace \, \e^{- \hat{H} } }{\tilde{V}^{-(\np + 2\nfree + \nbound)} \np! (\nfree!)^2 \nbound!},
\end{equation}
is expressed as a sum over the number paired ions $\nbound$, where $\nfree $ is constrained to $\nT - \nbound$ according to Eq.~\eqref{eq:nFree_constraint} and $\tilde{V} = V / \lambdaT^3$. In Eq.~\eqref{eq:partition_function_particle_picture}, we use a short-hand notation $\lbrace \dd \RR \rbrace \equiv \prod_{i=1}^{\np} \prod_{\alpha=1}^N \dd \RR_{i,\alpha}$ and $ \lbrace \dd \rr_{j} \rbrace \equiv \prod_{i=1}^{n_j} \dd \rr_{j,i}$ (for $j={\rm A,B,AB}$) to indicate that the integrals are performed over the positions of all polymer beads, solvated ions and bound ion pairs. 
The factorials in the denominator follow
from the fact that the polymers and the ${\rm A}^+$, ${\rm B}^-$ 
and ${\rm AB}$ solute are separately indistinguishable. Note that the particular combination $(\nfree!)^2 \nbound!$ implies that a bound state ${\rm AB}$ is distinguishable from any configuration of a free ${\rm A}^+$/${\rm B}^-$ pair. 

The logarithm of the partition function in Eq.~\eqref{eq:partition_function_particle_picture} can be used to define a free energy density $f(\rhob, \rhoT ; K_0, \lB, v, \dots) = - (\ln Z)/V$, where $V$ is the system volume, $\rhob = \np N /V$ is the bulk polymer bead number density and $\rhoT = \nT / V$ is the total bulk number density of A and B ions. We can compute the average number density of solvated ions $[{\rm A}^+]=[{\rm B}^-]=\langle \nfree \rangle / V$ through the first derivative of $f$ with respect to $K_0$, 
\begin{equation} \label{eq:rhofree_from_K0_derivative}
[{\rm A}^+] = - K_0 \frac{\partial f}{\partial K_0} . 
\end{equation}
The constraint \eqref{eq:nFree_constraint} then gives average number density of bound ion pairs as $[{\rm AB}] = \rhoT - [{\rm A}^+]$. Knowledge of $[{\rm A}^+]$ and $[{\rm AB}]$ can subsequently be used to compute the dissociation constant $K_{\rm d}$ according to Eq.~\eqref{eq:Kd_definition}. Other thermodynamic quantities of interest to this work are the polymer chain- and ion pair chemical potentials, $\mu_{\rm p} = N \partial f / \partial \rhob$ and $\mu_{\rm T} = \partial f / \partial \rhoT$ respectively, and the osmotic pressure $\Pi = \rhob \mu_{\rm p} / N + \rhoT \mu_{\rm T} - f$.

Note that $K_{\rm d}$ in Eq.~\eqref{eq:Kd_definition} in principle depends on how the bound state concentration $[{\rm AB}]$ is defined. In this work, we consider the contributing ${\rm AB}$ states as chemically distint (e.g.~through the appearance of a covalent bond between A and B, an ionic bond between
A$^+$ and B$^-$ with strength beyond that of simple electrostatic
attraction, or some other mechanism) from any configuration of free ${\rm A}^+$ and ${\rm B}^-$ molecules, as follows from the relation in Eq.~\eqref{eq:rhofree_from_K0_derivative} combined with the factorials in the denominator in Eq.~\eqref{eq:partition_function_particle_picture}. If $[{\rm AB}]$ is measured by exploiting the electrostatic properties of the ${\rm A}^+$--${\rm B}^-$ solution, rather than the chemical nature of the ${\rm AB}$ bond, a different definition of $[{\rm AB}]$ might be more appropriate where nearby ${\rm A}^+$ and ${\rm B}^-$ molecules are counted as effectively bound. This would be better captured by defining $[{\rm A}^+]$ and $[{\rm B}^-]$ through their corresponding activity parameters, similarly to how pH is formally defined through proton activity parameter rather than concentration\cite{Buck2002pHdef}. For simplicity,
this alternative approach is not pursued further in this work. 

\subsection{Deriving the field theory} \label{sec:deriving_field_theory}

The basis of the analytical calculations and lattice simulations is a field representation of the partition function in Eq.~\eqref{eq:partition_function_particle_picture}. To derive its corresponding field theory, we first decouple the interaction terms in $\hat{H}_1$ and $\hat{H}_2$ through standard Hubbard-Stratonovich transformations\cite{Edwards1965,Fredrickson2006}. This introduces two fields $w(\rr)$ and $\psi(\rr)$, conjugate to $\hat{\rho}(\rr)$ and $\hat{c}(\rr)$, respectively, and leads to
\begin{equation} \label{eq:Z_field_before_sum}
Z = \frac{\int \DD w \int \DD \psi \, Q_{\rm p}^{\np} \left( Q_{\rm A} Q_{\rm B} \right)^{\nT} \e^{-H_0[w,\psi]} \, S[w,\psi] }{\np! V^{-2\nT} \lambdaT^{3 \nT} K_0^{-\nT} Z_w Z_{\psi} }.
\end{equation}
up to an inconsequential multiplicative constant, where $H_0[w,\psi] = \int \dd \rr [ w^2 / 2 v + ({\bm \nabla} \psi)^2/ 8 \pi \lB ]$. The factor $S[w,\psi]$ contains the sum over $\nbound$,
\begin{equation} \label{eq:nbound_sum}
S = \sum_{\nbound=0}^{\nT} \frac{ 1 }{\nbound ! [(\nT - \nbound)!]^2} \left(  \frac{V^{-1} Q_{\rm AB} }{K_0 Q_{\rm A} Q_{\rm B}} \right)^{\nbound} .
\end{equation}
In the above expressions, $Q_i \equiv Q_i[w(\rr),\psi(\rr)]$ for $i={\rm p,A,B,AB}$ is the partition function for a type $i$ molecule in presence of chemical- and electrostatic potential potential fields $\ii \breve{w}(\rr) \equiv \Gamma \star \ii w(\rr)$ and $\ii\breve{\psi}(\rr) \equiv \Gamma \star \ii \psi(\rr)$ (where $\star$ denotes spatial convolution, i.e.~$\Gamma \star \phi(\rr) \equiv \int \dd \rr' \Gamma(\rr - \rr') \phi(\rr')$ for any function $\phi(\rr)$). For our model components, we have
\begin{equation}
\label{eq:polymer_Q}
Q_{\rm p} = \int \dd \RR_{1,\dots,N} \frac{ \e^{ -\frac{3}{2 b^2} \sum_{\alpha=1}^{N-1} {\Delta \RR_{\alpha}}^2 - \sum_{\alpha=1}^N \ii \breve{W}_{\alpha} } }{V (2 \pi b^2 / 3)^{3(N-1)/2}} , 
\end{equation}
where $\Delta \RR_{\alpha} \equiv \RR_{\alpha+1} - \RR_{\alpha}$, $\breve{W}_{\alpha} \equiv \breve{w}(\RR_{\alpha}) + \sigma_{\alpha} \breve{\psi}(\RR_{\alpha})$, and 
\begin{equation}
\label{eq:single_bead_Qs}
\begin{aligned}
 &Q_{\rm AB} = \frac{1}{V} \int \dd \rr \, \e^{-\gamma \ii \breve{w}(\rr) }, \quad Q_{\rm A} = \frac{1}{V} \int \dd \rr \, \e^{-\ii [\breve{w}(\rr) + \breve{\psi}(\rr) ]},  \\
& \text{and,} \qquad Q_{\rm B} = \frac{1}{V} \int \dd \rr \, \e^{-\ii [\breve{w}(\rr) - \breve{\psi}(\rr) ]}. 
\end{aligned}
\end{equation}
Note that all $Q_i$ are normalized to $Q_i[0,0] = 1$. The factors  $Z_w \equiv \int \DD w \exp(-\int \dd \rr w^2/2 v)$ and $Z_{\psi} \equiv \int \DD \psi \exp(- \int \dd \rr (\nabla \psi)^2/ 8 \pi \lB)$ have been inserted to provide additive density-independent contributions to the osmotic pressure that cancels the divergent contributions of the $| \kk | >a^{-1}$ modes of the functional integrals in Eq.~\eqref{eq:Z_field_before_sum}\cite{Villet2014,Lin2021}. 

In the thermodynamic limit, the sum over $\nbound$ in $S$ may be replaced by an integral that can be solved in the saddle-point approximation (this step is shown in Appendix \ref{sec:S_sum}). The resulting field picture representation of the partition function becomes
\begin{equation} \label{eq:partition_function_field_picture}
Z = \frac{\tilde{V}^{\np+\nT}}{\np! \nT!} \frac{\int \DD w \int \DD \psi \, \e^{-H[w,\psi] }}{ Z_w Z_{\psi} }   ,
\end{equation}
where the field Hamiltonian is
\begin{equation} \label{eq:field_hamiltonian}
H[w,\psi] = H_0 - \np \ln Q_{\rm p} - \nT \ln Q_{\rm AB} + \nT h_0(x) .
\end{equation}
Here, we have introduced a field operator $x[w,\psi]$, 
\begin{equation}\label{eq:x_def}
x[w,\psi] \equiv \frac{2 \rhoT}{K_0} \frac{Q_{\rm AB}}{Q_{\rm A} Q_{\rm B}}
\end{equation}
and the function
\begin{equation} \label{eq:h0_def}
h_0(x) = \ln\left(1 - \frac{\sqrt{1+2x} -1}{x} \right) - \frac{\sqrt{1+2x} -1}{x} . 
\end{equation}
In the following, we find it useful to define two additional functions, $h_1(x) \equiv x h_0'(x)$ and $h_2(x) \equiv x h_1'(x)$, where in particular
\begin{equation}
h_1(x) = \frac{\sqrt{1+2x} -1}{x} ,
\end{equation}
satisfies $0 \leq h_1(x) \leq 1$ for any real $x \geq 0$ (similarly, it may be shown that $h_2(x) \leq 0$). The physical interpretation of the field operator $h_1(x[w,\psi])$ becomes clear when applying the relation in Eq.~\eqref{eq:rhofree_from_K0_derivative} to the field partition function in Eq.~\eqref{eq:partition_function_field_picture}, leading to
\begin{equation}
\label{eq:[A]}
[ {\rm A}^+ ] =  \langle h_1(x) \rangle \rhoT , 
\end{equation}
i.e.~$h_1(x[w,\psi])$ is a field operator that averages to the fraction of dissociated ${\rm A}$\slash${\rm B}$ pairs. Correspondingly, $[ {\rm AB} ] = \rhoT \left[ 1 - \langle h_1(x) \rangle \right] $ gives the number density of the bound ion species. The field averaged $h_1(x[w,\psi])$ can therefore be used to evaluate $K_{\rm d}$ in Eq.~\eqref{eq:Kd_definition},
\begin{equation} \label{eq:Kd_field_theory}
K_{\rm d} = \rhoT \frac{\langle h_1 \rangle^2}{1 - \langle h_1 \rangle} . 
\end{equation}

\section{Analytical theory} \label{sec:analytical}

To understand the physical implications of our model in Eq.~\eqref{eq:partition_function_particle_picture}, we now proceed to approximately evaluate the functional integrals in the field representation of the partition function in Eq.~\eqref{eq:partition_function_field_picture}. The mean field theory (MFT) solution, which only accounts for the spatially homogeneous field configurations, captures the dominant effects on ion dissociation from excluded volume interactions and, in particular, showcases the crucial role of the parameter $\gamma$. However, effects from electrostatic interactions are entirely given by fluctuations of the charge-conjugate field $\psi(\rr)$, which are not accounted for in MFT\cite{PopovLeeFredrickson2007}. To capture the leading-order electrostatic effects, we next evaluate Eq.~\eqref{eq:partition_function_field_picture} in the random phase approximation (RPA). This accounts for Gaussian field fluctuations described by the expansion of $H[w,\psi]$ truncated beyond quadratic order in $w(\rr)$ and $\psi(\rr)$ (corresponding to fluctuations in density and charge, respectively). 

\subsection{Mean field theory } \label{sec:MFT}

A spatially homogeneous $\psi(\rr) = \bar{\psi}$ does not contribute in the field Hamiltonian in Eq.~\eqref{eq:field_hamiltonian} due to the over-all charge neutrality of the system, and hence does not contribute in MFT. The MFT solution for $w(\rr) = \bar{w}$, given by the vanishing first derivative of $H[w,0]$, satisfies
\begin{equation} \label{eq:MFT_w_solution} 
\ii \bar{w} = v \bar{\rho}_{\rm tot} ,
\end{equation}
where 
\begin{equation}
\bar{\rho}_{\rm tot} = \rhob + \left[ 2 \bar{h}_1 +  \gamma (1-\bar{h}_1) \right] \rhoT
\end{equation}
is the total concentration in MFT (counting each AB complex as $\gamma$ units), and $\bar{h}_1 \equiv h_1(x[\bar{w},0])$ with $x[\bar{w},0] = 2 \rhoT \e^{(2-\gamma) \ii \bar{w}} / K_0$. The exponential dependence of $\bar{h}_1$ on $\bar{w}$ means that Eq.~\eqref{eq:MFT_w_solution} generally lacks a closed-form solution, except for in the special case when $\gamma = 2$, and therefore has to be solved numerically for each given set of parameter values. 

Setting $\langle h_1 \rangle = \bar{h}_1$ in Eq.~\eqref{eq:Kd_field_theory} gives the MFT expression for the dissociation constant, which may be simplified to
\begin{equation}
\bar{K}_{\rm d} = K_0 \exp\left[ (\gamma-2) v \bar{\rho}_{\rm tot} \right] . 
\end{equation}
In the dilute limit, $\rhob, \rhoT \rightarrow 0 $, this expression reduces to $\bar{K}_{\rm d} = K_0$ verifying the claim in the previous section that the parameter $K_0$ is the dilute-limit MFT dissociation constant. In the opposite limit, where either $\rhob$ or $\rhoT$ are large, the dissociation constant instead either becomes exponentially suppressed (for $\gamma < 2$) or exponentially enhanced (for $\gamma > 2$) at densities $\gtrsim ( |\gamma-2| v)^{-1}$. Physically, this may be interpreted as the bound ${\rm AB}$ state being either favored or disfavored in a dense system depending on if it yields favorable excluded volume interactions compared to the dissociated state ${\rm A}^+$/${\rm B}^-$. At $\gamma=2$, the MFT dissociation constant is density independent. 

The MFT evaluation of the functional integrals ${Z_w}^{-1} {Z_{\psi}}^{-1} \int \DD w \int \DD \psi \, \exp(-H[w,\psi]) \approx \exp(-H[\bar{w},0])$ leads to the following expression for the free energy density,
\begin{equation*}
\bar{f} = - s_0 + \rhoT \bar{h}_0 + \frac{v}{2} \bar{\rho}_{\rm tot} \left[ \bar{\rho}_{\rm tot} - 2(2-\gamma) \bar{h}_1 \rhoT \right] , 
\end{equation*}
where $\bar{h}_0 \equiv h_0(x[\bar{w},0])$, and
\begin{equation*}
s_0 = \frac{1}{V} \ln \frac{V^{\np + \nT}}{\np! \nT!}  , 
\end{equation*}
is the configuration entropy density for a system with fully associated ions.
The MFT chemical potentials and osmotic pressure that follow are 
\begin{equation*}
\begin{aligned}
\bar{\mu}_{\rm p} &= \ln \frac{\rhob }{N} + v N  \bar{\rho}_{\rm tot} , \\
\bar{\mu}_{\rm T} &= \ln \left[ \rhoT (1-\bar{h}_1) \right] + v \gamma \bar{\rho}_{\rm tot} , \\
\bar{\Pi} &= \frac{\rhob}{N} + (1+\bar{h}_1) \rhoT + \frac{v}{2} \bar{\rho}_{\rm tot}^2. 
\end{aligned}
\end{equation*}

\subsection{Gaussian fluctuations:random phase approximation} \label{sec:RPA}
The Gaussian field fluctuations can be accounted for by expanding the field Hamiltonian $H[w,\psi]$ in Eq.~\eqref{eq:field_hamiltonian} to quadratic order about the MFT solution, and then performing the resulting Gaussian functional integrals in the partition function. This leads to the following correction term to the free energy density,
\begin{equation} \label{eq:RPA_free_energy}
f = \bar{f} + \frac{1}{2} \int \frac{\dd \kk}{(2 \pi)^3} \ln \left[ \frac{4 \pi \lB v}{\kk^2} \det \mathsf{G}(\kk) \right] ,
\end{equation}
where $\mathsf{G}(\kk)$ is the 2-by-2 matrix with the Fourier representation of the coefficients of the quadratic terms in the expansion of $H[w,\psi]$ and the factor $ 4 \pi \lB v / \kk^2$ comes from the product $Z_w Z_{\psi}$ in the denominator of Eq.~\eqref{eq:partition_function_field_picture}. Using the field basis $(w,\psi)$, we may write
\begin{equation*}
\mathsf{G}(\kk) = \mathsf{G}_0(\kk) + \rhob \mathsf{G}_{\rm p}(\kk) + \rhoT \mathsf{G}_{\rm T}(\kk) , 
\end{equation*}
where $\mathsf{G}_0(\kk) = \diag (1/v, \kk^2/4 \pi \lB)$, $\mathsf{G}_{\rm T}( {\bf 0}) =\diag( -(2-\gamma)^2 \bar{h}_2 , 0 )$, $\mathsf{G}_{\rm T}(\kk\neq {\bf 0}) = \hat{\Gamma}(\kk)^2 \diag( \gamma^2 + (2-\gamma^2)\bar{h}_1, 2 \bar{h}_1 )$, $\mathsf{G}_{\rm p}( {\bf 0}) =\diag( 0, 0 )$ and
\begin{equation*}
\mathsf{G}_{\rm p}(\kk\neq {\bf 0}) = \hat{\Gamma}(\kk)^2 \begin{pmatrix}
g_{\rm mm}(\kk) & g_{\rm mc}(\kk) \\
g_{\rm mc}(\kk) & g_{\rm cc}(\kk)
\end{pmatrix}. 
\end{equation*}
The entries of $\mathsf{G}_{\rm p}(\kk\neq {\bf 0})$ are the standard single chain density-density-, density-charge- and charge-charge correlations following from the expansion of $Q_{\rm p}$, i.e.~$g_{\rm mm} (\kk)= \sum_{\alpha,\beta=1}^N \e^{- |\alpha - \beta| b^2 \kk^2 / 6} / N$, $g_{\rm mc}(\kk) = \sum_{\alpha,\beta=1}^N \sigma_{\alpha}  \e^{- |\alpha - \beta| b^2 \kk^2 / 6} / N$ and $g_{\rm cc}(\kk) = \sum_{\alpha,\beta=1}^N \sigma_{\alpha} \sigma_{\beta} \e^{- |\alpha - \beta | b^2 \kk^2 / 6} / N$. In the above expressions, $\hat{\Gamma}(\kk) = \e^{- a^2 \kk^2 / 2}$ is the Fourier transformation of the Gaussian smearing function $\Gamma(\rr)$. 

The derivatives of Eq.~\eqref{eq:RPA_free_energy} with respect to species numbers, volume and $K_{0}$ yields corrections from Gaussian field fluctuations to the chemical potentials, osmotic pressure and fraction of dissociated ion pairs, $\mu_{\rm p} = \bar{\mu}_{\rm p} + \mu_{\rm p}^{({\rm fl})}$, $\mu_{\rm p} = \bar{\mu}_{\rm p} + \mu_{\rm p}^{({\rm fl})}$, $\mu_{\rm T} = \bar{\mu}_{\rm T} + \mu_{\rm T} ^{({\rm fl})}$, $\Pi = \bar{\Pi} + \Pi^{({\rm fl})}$ and $\langle h_1(x) \rangle = \bar{h}_1 + h_1^{({\rm fl})}$, respectively. The full expressions for the RPA corrections $X^{({\rm fl})}$ are given in Appendix \ref{sec:RPA_expressions}. 

The RPA contributions to our thermodynamic observables of interest all involve integrals over wave numbers $\kk$ that generally need to be computed numerically. A special case, that can be treated fully analytically, is the dilute limit of $h_1^{({\rm fl})}$, leading to the following RPA expression for the dilute limit dissociation constant,
\begin{equation}
\lim_{\rhob, \rhoT \rightarrow 0} K_{\rm d} = K_0 \left( 1 + \frac{(2-\gamma^2) v}{16 \pi^{3/2} a^3} + \frac{\lB}{\sqrt{\pi} a} \right)^{-1} . 
\end{equation}
This shows that the bare parameter $K_0$ can no longer be interpreted as the dilute limit dissociation constant when fluctuations are included in the calculation. 

\section{Field Theory Simulations} \label{sec:FTS}

We complement our findings from RPA calculations with field theoretical simulations (FTS) that fully capture the fluctuations of the fields $w(\rr)$ and $\psi(\rr)$. Other alternative simulation approaches to biomolecular LLPS include explicit chain MD simulations\cite{Silmore2017,Dignon2018,DasAminLinChan2018,SumanJPCB2018} and finite-size scaling theory\cite{NilssonIrback2020,NilssonIrback2021}. In FTS, equilibrium evolution of the system dictated by the Hamiltonian Eq.~\eqref{eq:field_hamiltonian} is studied by following a complex Langevin (CL) prescription \cite{FredricksonRev2002, Fredrickson2006, ParisiWu1981, Parisi1983, Klauder1983,ChanHalpern1986} where the real-valued continuous fields $w$ and $\psi$ are analytically continued to their respective complex planes and evolved in the fictitious CL time through the equations given by
\begin{equation}
\label{eq:CL_equations}
\begin{aligned}
\frac{\partial w(\rr, t)}{\partial t} =& - \left[ \ii \tilde{\rho}(\rr, t) +\frac{w(\rr, t)}{v}  \right] + \eta_w(\rr, t),\\
\frac{\partial \psi(\rr, t)}{\partial t} =& - \left[ \ii \tilde{c}(\rr, t) - \frac{\bm{\nabla}^2 \psi(\rr, t)}{4 \pi \lB} \right] + \eta_{\psi}(\rr, t).
\end{aligned}
\end{equation}
This allows ensemble averages to be computed as asymptotic CL time averages. In Eq.~\eqref{eq:CL_equations}, $\eta_{w}$ and $\eta_{\psi}$ are real valued random numbers with zero mean and variance $2\delta (\rr - \rr') \delta (t - t')$. The field operators for the number- and charge densities, $\tilde{\rho} \equiv \tilde{\rho}_{\rm b}  + \tilde{\rho}_{\rm A} + \tilde{\rho}_{\rm B} + \tilde{\rho}_{\rm AB}$ and $\tilde{c} \equiv \tilde{c}_{\rm b} + \tilde{\rho}_{\rm A} - \tilde{\rho}_{\rm B}$, respectively, are obtained from
\begin{equation}
\label{eq:density_operators_polymers}
\tilde{\rho}_{\rm b}(\rr, t) = \ii \np \frac{\delta \ln Q_{\rm p}}{\delta w(\rr, t)} ,\text{ and } \tilde{c}_{\rm b}(\rr, t) = \ii\np \frac{\delta \ln Q_{\rm p}}{\delta \psi(\rr, t) } ,
\end{equation}
for polymers, and
\begin{equation}
\label{eq:density_operators_ions}
\begin{aligned}
\tilde{\rho}_{i}(\rr, t) &= h_1(x) \ii \nT \frac{\delta \ln Q_{i} }{\delta w(\rr, t)} , \quad i={\rm A,B}, \\
\tilde{\rho}_{\rm AB}(\rr, t) &= \left[ 1-h_1(x) \right] \ii \nT \frac{\delta \ln Q_{\rm AB}}{\delta w(\rr, t)} ,
\end{aligned}
\end{equation}
for ions where $x[w,\psi]$ is defined in Eq.~\eqref{eq:x_def}. Detailed expressions for the above density operators can be found in Appendix \ref{sec:appn_fts}. Note that contributions from field fluctuations up to all orders are kept in Eqs.~\eqref{eq:density_operators_polymers} and \eqref{eq:density_operators_ions}. We numerically solve Eq.~\eqref{eq:CL_equations} on a cubic lattice with periodic boundary conditions and lattice spacing $\Delta x$, using a semi-implicit first order time stepping method\cite{Lennon2008SI,Lin2021} with a CL time-step $\Delta t=10^{-3}b^3$. Use of the semi-implicit time stepping method results in significantly better numerical stability compared to the Euler--Maruyama type explicit time-discretization methods \cite{Lennon2008SI}. As with the CL-time discretization, the spatial discretization of the continuous fields $w$ and $\psi$ is an approximation, and, formally, FTS only exactly reproduces the continuum field theory in the limit $\Delta x \rightarrow 0$. However, the Gaussian smearing already provides a strong exponential suppression of contributions from field fluctuations on distance scales smaller than the smearing length $a$, such that these modes may be omitted with a negligible numerical effect on physical observables. In this work, we therefore set $\Delta x = a$ in all FTS computations.

Thermally averaged bulk densities of solvated and bound ion concentrations can be obtained from the field averaged value of $h_1(x)$. Information about how the components are spatially distributed in the system can be gleaned from potential of mean forces (PMFs). The PMF $U_{i,j}(r)$ between two components $i$ and $j$ describe the free-energy landscape for the separation $r$ between two units of $i$ and $j$, and is related to the corresponding radial distribution function (RDF) $g_{i,j}(r)$ through the relation
\begin{equation}
\label{eq:PMF_defn}
U_{i,j}(r) = - \ln g_{i,j}(r) .
\end{equation}
Here, $U_{i,j}(r)$ is given in units of the thermal energy $k_{\rm B} T$. Since explicit particle coordinates have been traded off to the field degrees of freedom in the field picture, RDFs have to be computed through their field operators defined by \cite{PRE2021Pal}
\begin{widetext}
\begin{equation}
\label{eq:RDFs}
\begin{aligned}
g_{\rm p,p}(|\rr-\rr'|) =& \frac{1}{\rhob \rhob' }\bigg[ \frac{\ii}{v}\left\langle \tilde{\rho}_{\rm b}(\rr) w(\rr') \right\rangle - \sum_{i}\left\langle \tilde{\rho}_{\rm b}(\rr) \tilde{\rho}_{i}(\rr') \right\rangle \bigg] - \frac{1}{\rhob'} \frac{\e^{-(\rr-\rr')^2/4a^2}}{(4\pi a^2)^{3/2}},\quad i = \text{A, B, AB},\\
g_{\rm p, S}(|\rr-\rr'|) =& \frac{\left\langle \tilde{\rho}_{\rm b}(\rr) \left[ \tilde{\rho}_{\rm A}(\rr') + \tilde{\rho}_{\rm B}(\rr') \right] \right\rangle}{2\rho_{\rm b}\rho_{\rm T}\langle h_1\rangle} , \quad\text{and},\quad
g_{\rm p, B}(|\rr-\rr'|) = \frac{\left\langle \tilde{\rho}_{\rm b}(\rr) \tilde{\rho}_{\rm AB}(\rr') \right\rangle}{\gamma \rhob \rho_{\rm T}(1-\langle h_1\rangle)},
\end{aligned}
\end{equation}
\end{widetext}
where the subscripts p, S and B on the RDFs stand for polymer bead, solvated and bound, respectively. The last term in the expression for 
$g_{\rm p,p}(r)$ subtracts the contribution from the polymer bead with itself. The RDFs in Eq.~\eqref{eq:RDFs} have been normalized to $g_{i,j}=1$ when units of $i$ and $j$ are uncorrelated, which is expected e.g.~at large $r$ if the system contains a single liquid phase. In Eq.~\eqref{eq:RDFs}, the factor $\rhob'=\rhob - N/V$ provides the correct finite-volume correction to the polyampholyte--polyampholyte RDF normalisation, and approaches $\rhob$ at large $V$. 

If the system is in a globally inhomogeneous state, e.g.~by containing several co-existing macro-phases, certain PMFs may approach non-zero values at large separations $r$. In particular, the large $r$ behavior of the polyampholyte--polyampholyte PMF $U_{\rm p,p}(r)$ indicates if the polymers are homogeneously distributed on large scales or are concentrated in e.g.~a dense droplet. The information of the influences of the solvated ions and the neutral ion-pairs on phase separation, on the other hand, are obtained from the polyampholyte--solvated-ions and polyampholyte--ion-pair PMFs, respectively.

\section{Results} \label{sec:results}
\begin{figure}
\includegraphics{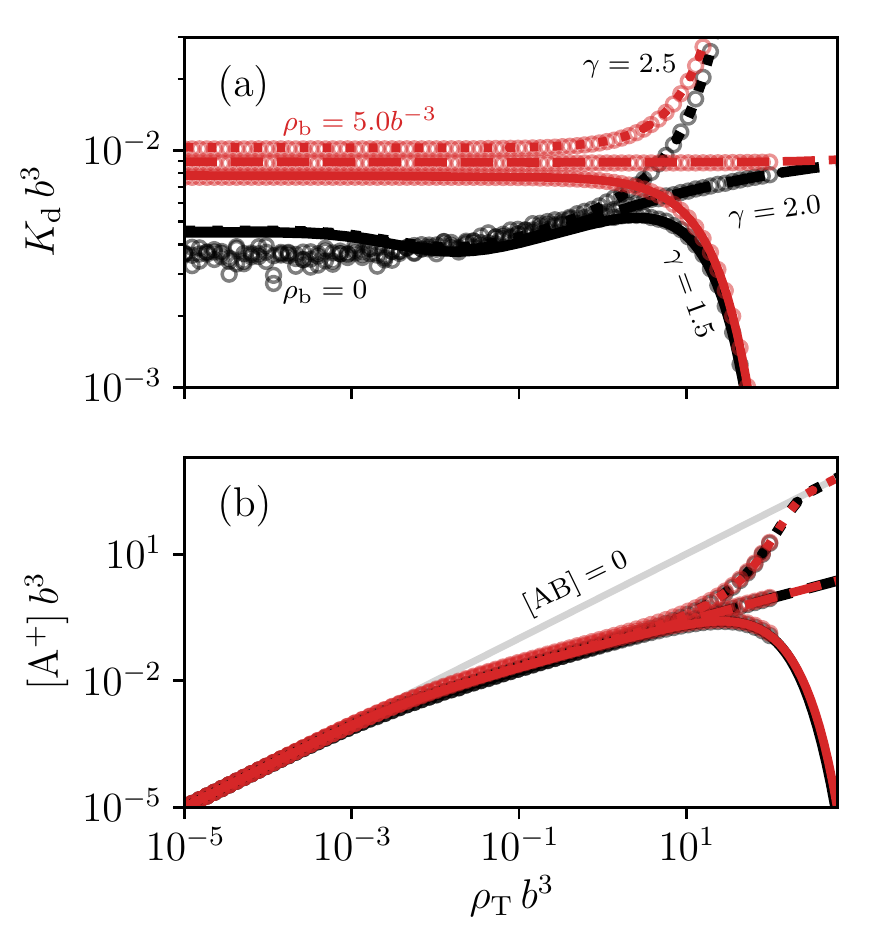}
\caption{The dissociation constant $K_{\rm d}$ (a) and solvated ion density $[{\rm A}^+]= [{\rm B}^-]$ (b) computed in RPA (curves) and FTS (circles) as functions of $\rhoT = [{\rm A}^+] + [{\rm AB}]$. Black and red correspond to systems with zero and high polymer densitiy ($\rhob=0$ and $\rhob=5b^{-3}$), respectively. Values of $\gamma$ are indicated by the linestyles with $\gamma=1.5$, $2$ and $2.5$ corresponding to solid, dashed and dotted lines, respectively. The gray line in (b) indicates where all ions exist in the solvated state, i.e.~$ [{\rm A}^+]=\rhoT$ and $ [{\rm AB}]=0$. All results are computed using the sv15 charge sequence in Fig.~\ref{fig:cartoons}(b) with $\lB=0.6 b$, $v=0.05b^3$, $K_0=0.01b^{-3}$ and $a=b/\sqrt{6}$. The FTS results were computed on $32^3$ cubic lattice with lattice spacing $\Delta x = b/\sqrt{6}$ and a Complex-Langevin time-step $\Delta t = 10^{-3} b^3$. } \label{fig:Kd_rhoA}
\end{figure}

The effects on ion pairing from the solution condition can be understood from Fig.~\ref{fig:Kd_rhoA}, showing $K_{\rm d}$ and $[{\rm A}^+] =[{\rm B}^-]$ computed in RPA and FTS as functions of $\rhoT$ at $\gamma=1.5$, $2.0$ and $2.5$ in polymer empty ($\rhob=0$) and dense ($\rhob=5b^{-3}$) systems with polyampholyte species sv15. The FTS calculations were done on a $32^3$ lattice. The plots in Fig.~\ref{fig:Kd_rhoA} show excellent agreement between RPA and FTS except in the very dilute limit where field fluctuations beyond Gaussian order are expected to become important. Proper sampling of important field configurations may not be possible through CL methods when the system is dictated by relatively high excluded volume interaction together with short range hydrophobicity type interactions\cite{Nilsson2022}. The agreement between the RPA and FTS results seen here is thus reassuring that $\Delta t$ and $\Delta x$ values used in the FTS implementation provide sufficient numerical accuracy. When the total density is $\lesssim (v | 2-\gamma |)^{-1}$, the dominant effect on the dissociation constant $K_{\rm d}$ comes from charges in the surroundings (either free ions or charged polymer residues) that screen out the electrostatic component of the A-B binding energy, thus increasing the value of $K_{\rm d}$. In Fig.~\ref{fig:Kd_rhoA} (a), this effect underlies both the slow increase of $K_{\rm d}$ with increasing $\rhoT$ at $\rhob=0$, and difference between the $\rhob=0$ and $\rhob=5b^3$ curves. At higher densities, where effects from excluded volume interactions instead dominate, the exponential dampening (for $\gamma<2$) or enhancement (for $\gamma>2$) of $K_{\rm d}$ determines whether mainly bound states ${\rm AB}$, or free ${\rm A}^+$ and ${\rm B}^-$ are present in the system, as can be seen in Fig.~\ref{fig:Kd_rhoA} (b). 

\begin{figure*}
\includegraphics{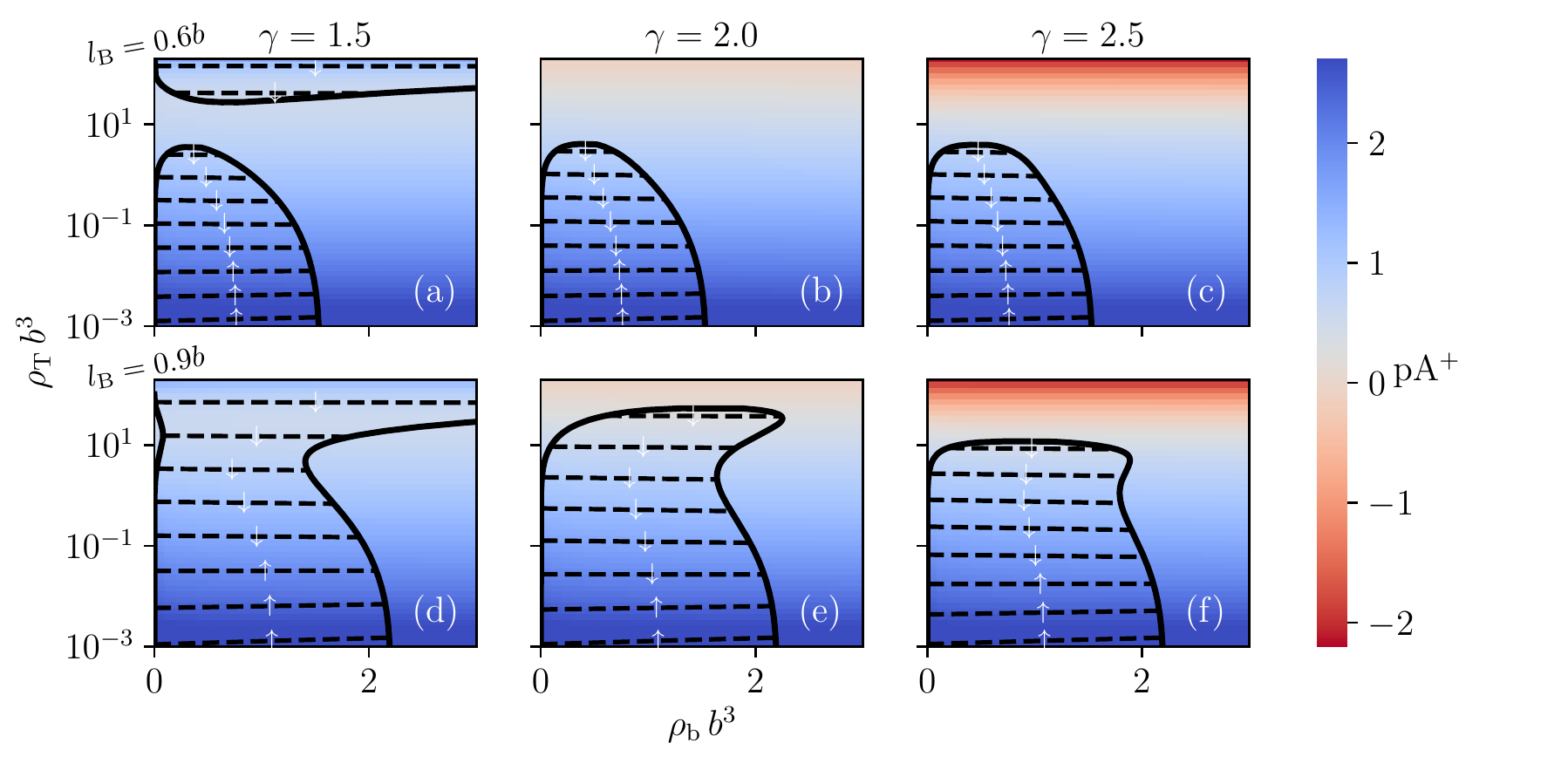}
\caption{RPA phase diagrams in the $(\rhob,\rhoT)$ plane. Sub-figures (a--c) in the top row correspond to $\gamma=1.5$, $2$ and $2.1$, respectively, and all with $\lB=0.6b$. In the bottom row, sub-figures (d--f) are for a common $\lB=0.9b$ and $\gamma$ values $1.5$, $2$ and $2.1$, respectively. Other parameter values are $v=0.05b^3$, $K_0=0.01b^{-3}$ and $a=b/\sqrt{6}$. The color gradient shows the density of solvated ions $[{\rm A}^+]$, expressed through Eq.~\eqref{eq:pA}. Tie-lines indicated by $\uparrow$ and $\downarrow$ have positive and negative slopes, respectively.} \label{fig:phase_diags_gamma}
\end{figure*}

Fig.~\ref{fig:phase_diags_gamma} shows RPA phase diagrams for the system with sv15 polyampholyte species computed by matching the osmotic pressure $\Pi(\rhob,\rhoT)$ and chemical potentials $\mu_{\rm p}(\rhob, \rhoT)$ and $\mu_{\rm T}(\rhob, \rhoT)$ along the phase boundary\cite{Kardar2007,Lin2021}. The binodal curves (solid black curves) enclose the co-existence regions, where the co-existing bulk density values of $\rhob$ and $\rhoT$ are connected by tie-lines (dashed lines).
The background color gradient shows the density of solvated ions, $[{\rm A}^+]$, expressed through
\begin{equation} \label{eq:pA}
{\rm pA}^+ \equiv - \log_{10}[{\rm A}^+]b^3 \, . 
\end{equation}
The phase diagrams in Fig.~\ref{fig:phase_diags_gamma} display strong dependence on $\lB$ and $\gamma$, and, in particular, predict the possibility of re-entrant phase separation with increasing $\rhoT$. A high $\rhoT$ re-entrant phase separation region occurs at $\lB=0.6 b$ for $\gamma=1.5$, but vanishes for $\gamma \geq 2$, which can be understood from the preferred ion-pairing state in the dense system. As can be seen from Fig.~\ref{fig:Kd_rhoA}, for $\gamma=1.5$, a large $\rhoT$ strongly favors the charge neutral bound state AB which only interacts through excluded volume repulsion. The resulting crowding effects leads to an effective attraction between the chains that further promotes phase separation which enables the re-entrant phase separation. At $\gamma=2.5$, the high $\rhoT$ system contains almost exclusively solvated ions that strongly screen out the polymer electrostatic interactions, thus inhibiting phase separation. At the boundary value $\gamma=2$, although the high $\rhoT$ state still contains a substantial amount of bound ions, the small amount of solvated ions (behaving roughly as $[A^+] \sim \sqrt{\rhoT}$) is enough to dissolve the condensates. When increasing $\lB$, the two disconnected co-existence regions at $\gamma=1.5$, merge into one.  Note that the considered polyampholyte species phase separates for both the $\lB$ values used in Fig.~\ref{fig:phase_diags_gamma} in the absence of any ions.

\begin{figure}
\includegraphics{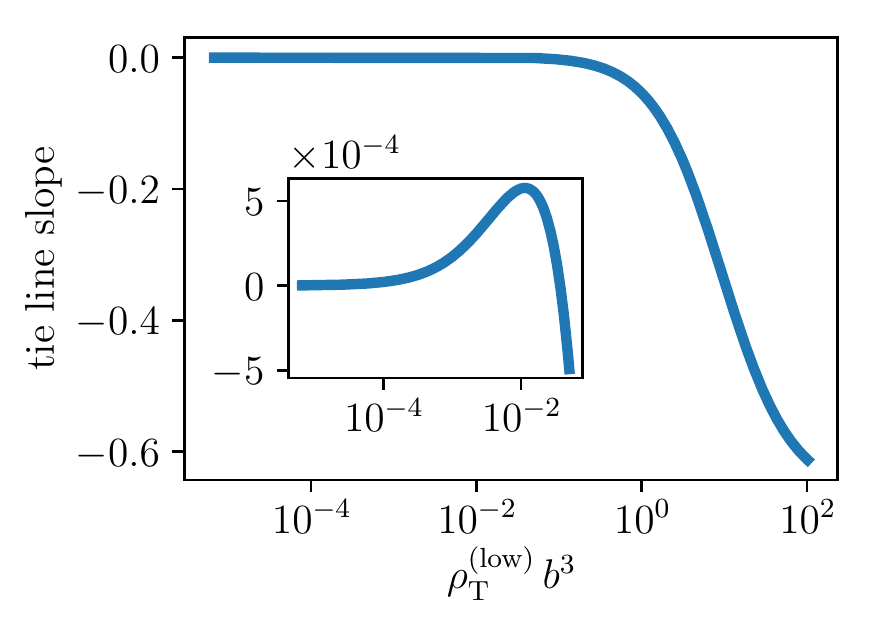}
\caption{Slopes of the tielines for the $\gamma=1.5$ and $\lB=0.9b$ phase diagram in Fig.~\ref{fig:phase_diags_gamma} (d), as function of the $\rhoT$ value of the polymer-dilute phase. Inset shows the region where tie-line slope changes sign.} \label{fig:tie_line_slopes}
\end{figure}
All tie-lines in Fig.~\ref{fig:phase_diags_gamma} are roughly horizontal, but tend to have slightly positive and negative slopes (indicated by $\uparrow$'s and $\downarrow$'s) when the ions dominantly are solvated or exist in the bound state, respectively. This is exemplified in Fig.~\ref{fig:tie_line_slopes}, where we show the tie-line slopes for the phase diagram in Fig.~\ref{fig:phase_diags_gamma} (d) (i.e.~at $\lB=0.9b$ for $\gamma=1.5$). The inset of Fig.~\ref{fig:tie_line_slopes} shows the region where the tie-line slopes change sign, beyond which the tie-line slope becomes increasingly more negative at higher $\rho_{\mathrm{T}}$ values.

\begin{figure}
\includegraphics{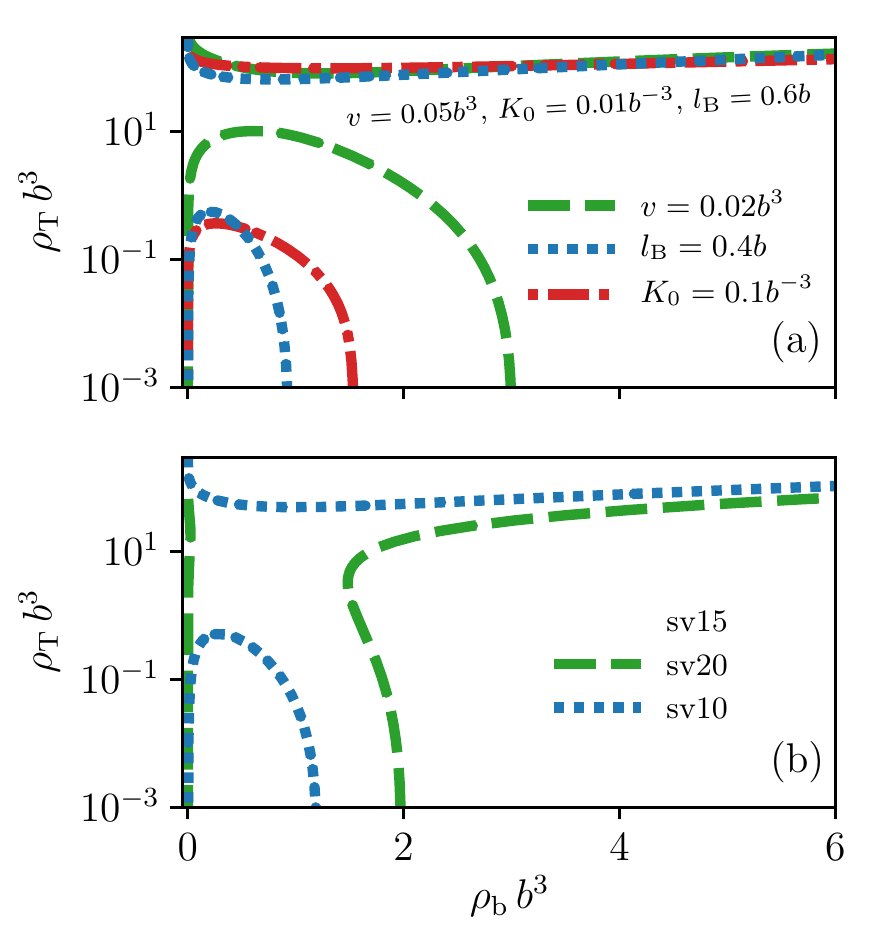}
\caption{Re-entrant phase behaviour dependencies on (a) model parameters and (b) polyampholyte charge sequece. The black curve in (a) and (b) correspond to the reference sv15 phase diagram for $\gamma=1.5$ and $\lB=0.6b$ in Fig.~\ref{fig:phase_diags_gamma} (top left). Other curves in (a) correspond to varying either $v$, $\lB$ and $K_0$. Other curves in (b) instead show the charge sequence dependence (sequences used are shown in Fig.~\ref{fig:cartoons}).} \label{fig:phase_diags_lB_v_seq}
\end{figure}

In Fig.~\ref{fig:phase_diags_lB_v_seq}, we focus on the $\gamma=1.5$ and $\lB=0.6b$ phase diagram of Fig.~\ref{fig:phase_diags_gamma} and investigate the dependence of the re-entrant phase behavior on the model parameters and the polymer charge sequence. The excluded volume parameter $v$ plays different roles in the upper and lower regions of the re-entrance phase diagrams, as seen in Fig.~\ref{fig:phase_diags_lB_v_seq} (a). While reducing the value of $\lB$ (which reduces the strength of electrostatic interactions) or increasing $K_0$ (giving more solvated ions that provide electrostatic screening) both decrease the phase separation propensity by shrinking the two co-existence regions, a reduced $v$ simultaneously enlarges the low $\rhoT$ region and shrinks the high $\rhoT$ region. The excluded volume interactions with ${\rm AB}$ molecules therefore act as to stabilize the high $\rhoT$ condensates, while the excluded volume interactions among the polymers in the low $\rhoT$ region inhibit phase separation. In Fig.~\ref{fig:phase_diags_lB_v_seq} (b), we instead swap the sv15 charge sequence by either sv10 or sv20, which are characterized by smaller or larger blocks of consecutive same-sign charges, respectively. This degree of ``blockiness'' can be quantified e.g.~by the $\kappa$ parameter of Das and Pappu\cite{DasPappu2013}, or by the sequence charge decoration parameter of Sawle and Gosh \cite{SawleGosh2015}, and has been shown to strongly correlate with phase separation propensity\cite{LinPRL,SumanJPCB2018,DasAminLinChan2018,McCartyDelaneyDanielsenFredricksonShea2019,Danielsen2019}.
Fig.~\ref{fig:phase_diags_lB_v_seq} (b) shows that the phase diagram exhibits a strong dependence on the charge sequence blockiness, and that even the topology of the co-existence region may depend on the sequence (c.f.~two disconnected co-existence regions of sv10 and sv15 versus the connected sv20 co-existence region).

\begin{figure*}
\includegraphics{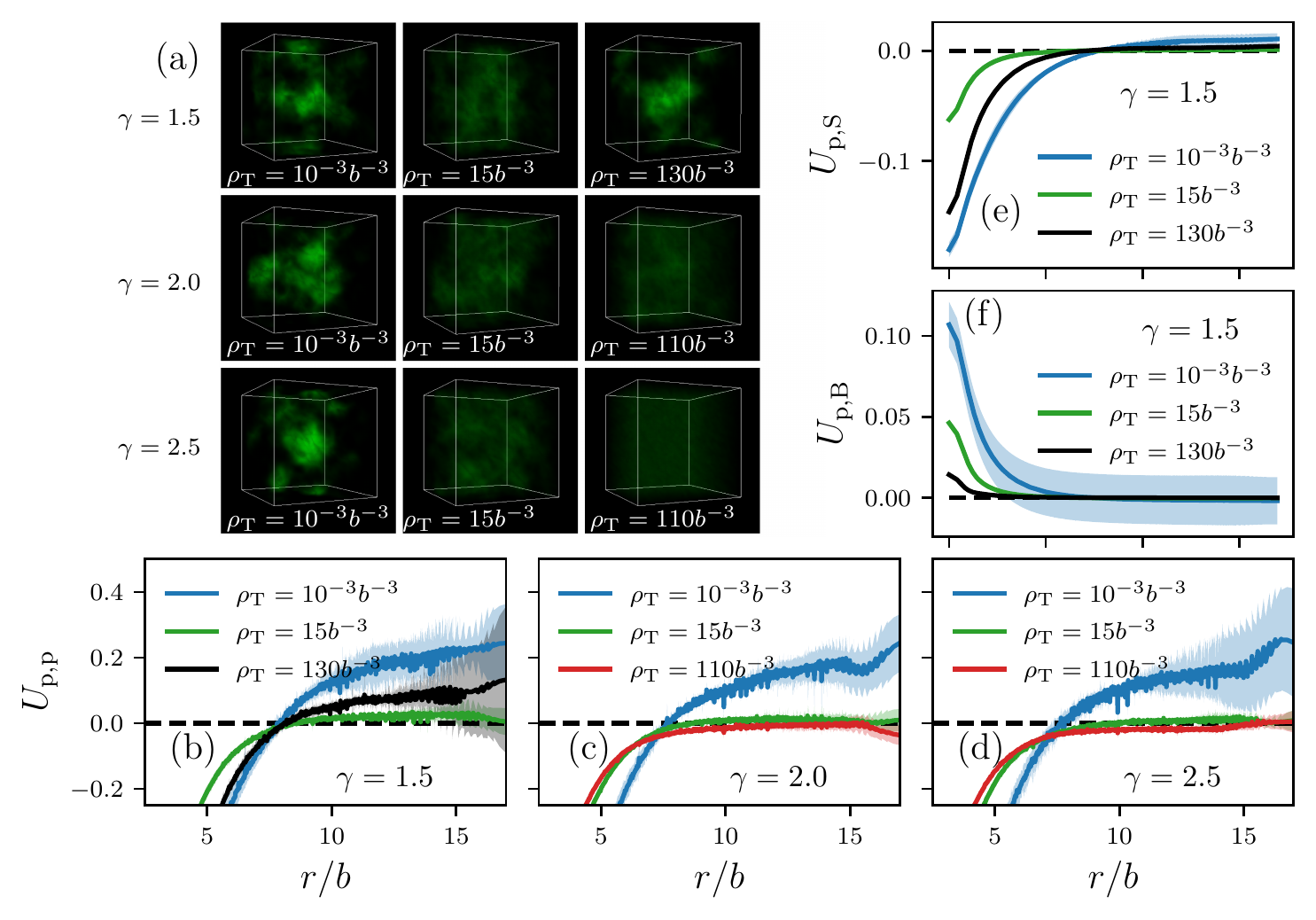}
\caption{FTS results for the species sv15 at $\lB=0.6b$, $\rho_{\rm b}=0.1b^{-3}$, $K_0=0.01b^{-3}$and $v=0.05b^3$. (a) Representative snapshots of real non-negative parts of polymer bead density operator $\tilde{\rho}_{\rm b}$ at various total bulk ion-densities. (b-d) Polymer-Polymer PMFs at $\gamma=1.5, 2$ and $2.5$. (e) and (f) Polymer--Solvated-ions PMFs and Polymer--ion-pairs PMFs, respetively, at $\gamma=1.5$.  The solid lines and the shaded regions around the solid lines are mean values and standard deviations of the respective observables computed over five independent simulations, respectively.} \label{fig:FTS_results}
\end{figure*}
We qualitatively verify some of the LLPS properties obtained from RPA calculations using FTS calculations of the PMFs $U_{\rm p,p}(r)$, $U_{\rm p,S}(r)$ and $U_{\rm p,B}(r)$. For the PMF calculations, we now use a larger ($48^3$) lattice and focus on the polymer charge sequence sv15. Other fixed parameter values are $\lB=0.6b^{-3}$, $v=0.05b^3$, $K_0=0.01b^{-3}$ and $\rho_{\rm b}=0.1b^{-3}$. We simulate the system at various initial reactant concentrations and values of $\gamma$. The results for different $\rho_{\rm T}$ values are shown in Fig.~\ref{fig:FTS_results}. To obtain reliable statistics, we ran 5 independent simulations for each $\rho_{\rm T}$. For each independent run, out of a total $1.5\times10^{5}$ CL time-steps, we discard the first $5\times10^4$ steps and use the rest to compute field averages of the RDFs using a sampling interval of $500$ steps. The equilibration time was taken as the CL time step after which the real part of the single unit partition functions stabilized around constant values and the production phase of the simulations were run until a reasonably small standard deviation of the computed observables, $K_{\mathrm{d}}$ and PMFs, were reached. In Fig.~\ref{fig:FTS_results}a, we show representative snapshots of the field picture polymer bead density $\tilde{\rho}_{\rm b}(\rr)$, defined in Eq.~\eqref{eq:density_operators_polymers}, at different reactant densities for $\gamma=1.5, 2$ and $2.5$. Consistent with the RPA phase diagrams of Fig.~\ref{fig:phase_diags_gamma}, we see polymer droplet formation for all three $\gamma$ values at relatively low reactant concentrations, but at high reactant concentrations, droplet formation happens only at $\gamma=1.5$. Different LLPS behavior is also evident from the polymer-polymer PMF plots shown in Figs.~\ref{fig:FTS_results}b-d where at the highest reactant concentrations shown, the PMFs go to zero for $\gamma=2$ and $2.5$ and to a small but non-trivial positive value for $\gamma=1.5$ at large separations. The short-range effective attraction between polymers and solvated ions or repulsion between polymers and bound ${\rm AB}$ molecules, respectively, demonstrated by the corresponding PMFs in Fig.~\ref{fig:FTS_results}e-f, are consistent with the tie-line slopes (see Fig.~\ref{fig:tie_line_slopes}) which tend to be positive and negative when ions are in their solvated and bound states, respectively.

\section{Conclusions} \label{sec:conclusions}

In this work, we have introduced a model for polyampholytes undergoing LLPS in presence of monovalent reactants ${\rm A}^+$ and ${\rm B}^-$ that can form a chemically distinct bound state ${\rm AB}$. The dissociated ions have diametrically different effects on the polymer phase separation from their electrically neutral bound counterpart: While the free ions screen the LLPS driving electrostatic interactions between the chains (thus decreasing the phase separation propensity), the bound pairs ${\rm AB}$ instead function as a crowding agent that promotes LLPS. Conversely, the crowded and highly charged environment inside the polyampholyte condensates have non-trivial effects on the ${\rm A-B}$ dissociation constant compared to the co-existing dilute phase. This complex interplay between polymer phase separation and ion dissociation is studied using both analytical and simulation approaches which, in particular, consistently point towards a novel mechanism for re-entrant phase behavior under the circumstances where bound ion pairs yield favorable excluded volume interactions over their dissociated counterpart. 

Among the plethora of biological functions of IDP condensates that are currently being uncovered, regulating chemical reactions seem to be one of their major functional roles \cite{AlbertiGladfelterMittag2019}. Additionally, chemical reactions in the cellular environment have been shown to be able to both dissolve condensates and trigger their formation \cite{Wippich2013,Zwicker2015,Zwicker2017,Wurtz2018,Kirschbaum2021}. The field theoretic approach presented in this work constitutes a major methodological advancement for modelling of such phenomena, due to our treatment of fluctuations compared to existing MFT theories for polymer phase separation in chemically reactive environments. Accounting for field fluctuations is necessary to capture amino-acid sequence dependence\cite{LinPRL}, and is essential for describing phase separation driven by the electrostatic interactions\cite{PopovLeeFredrickson2007} characterized by many IDP species. 

The proposed mechanism for ion-triggered re-entrant phase separation relies on electrostatic screening from solvated ions ${\rm A}^+$/${\rm B}^-$ combined with crowding effects from their electrically neutral bound state ${\rm AB}$. While Coulombic screening is a long established consequence from free ions, LLPS promoted by molecular crowding is a relatively less so studied phenomenon\cite{ParkFredrickson2020,JonasJPCB2021,ChanDill1997,Ray1971}. Our model connects these two effects through ion association, quantified by the bare dissociation constant $K_0$. Although it remains to be seen if this particular mechanism for re-entrance is realized in nature, we nevertheless believe that our framework for including chemical reactions in RPA and FTS will be applicable to a wide range of systems through minor phenomenological modifications.

\begin{acknowledgments}

The authors thank Yi-Hsuan Lin and Suman Das for insightful discussions.
This work was supported by Canadian Institutes of Health Research grant 
NJT-155930 and Natural Sciences and Engineering Research Council of Canada 
grant RGPIN-2018-04351 as well as computational resources provided 
generously by Compute/Calcul Canada. The data that support the findings of this study are available from the corresponding authors upon reasonable request. J.W.~and T.P.~contributed equally to this work.
\end{acknowledgments}

\appendix

\section{Calculation of the sum over $\nbound$ in deriving the field Hamiltonian} \label{sec:S_sum}
Here, we show how the sum $S[w,\psi]$ over bound ion pairs $\nbound$ in Eq.~\eqref{eq:nbound_sum} is computed in the thermodynamic limit. We express $S[w,\psi]$ as
\begin{equation*}
S = \sum_{\nbound = 0}^{\nT} \frac{ y^{\nbound} }{\nbound! [(\nT - \nbound)!]^2}, \quad y \equiv \frac{ Q_{\rm AB} }{ K_0 V Q_{\rm A} Q_{\rm B} }. 
\end{equation*}
In the thermodynamic limit, we can replace the sum by an integral which we solve in the saddle-point approximation,
\begin{equation*}
S = \int_0^{\nT} \dd \nbound \e^{-I(\nbound)} = \e^{-I(\nbound^*)},
\end{equation*}
where
\begin{equation*}
I(\nbound) = \ln\nbound! + 2 \ln(\nT - \nbound)! - \nbound \ln y, 
\end{equation*}
and $\nbound^*$ satisfies the saddle condition $I'(\nbound^*)=0$. Using Stirling's approximation $\ln n! \approx n \ln n - n$ on the logarithm of the factorials gives the saddle condition
\begin{equation*}
y = \frac{\nbound^* }{(\nT - \nbound^*)^2} , 
\end{equation*}
which has two solutions,
\begin{equation*}
\nbound^* = \nT \left[ 1 + \frac{1 \pm \sqrt{1 + 2 x} }{x} \right], \quad x=2\nT y. 
\end{equation*}
However, only the `$-$' solution is contained in the integration interval $0 \leq \nbound \leq \nT$ and contributes to $S[w,\psi]$. Plugging this solution into $I(\nbound^*)$ gives
\begin{equation*}
I(\nbound^*) = \ln \nT! - \nT \ln y + \nT h_0(x), 
\end{equation*}
where we again have used Stirling's approximation in writing $\nT \ln \nT - \nT \approx \ln \nT!$. The function $h_0(x)$ is defined in Eq.~\eqref{eq:h0_def}. The thermodynamic limit of $S[w,\psi]$ then becomes
\begin{equation*}
S[w,\psi] = \frac{1}{\nT!} \left[  \frac{ Q_{\rm AB} }{ K_0 V Q_{\rm A} Q_{\rm B} } \right]^{\nT} \e^{-\nT h_0(x) } . 
\end{equation*}
Plugging this expression into Eq.~\eqref{eq:Z_field_before_sum} gives the final field representation of the partition function in Eq.~\eqref{eq:partition_function_field_picture}. 

\section{Complete RPA expressions for $\mu_{\rm p,T}$, $\Pi$ and $K_{\rm d}$} \label{sec:RPA_expressions}
This appendix describes the derivation of the RPA expressions for the chemical potentials, osmotic pressure and dissociation constant. We start by expanding the field Hamiltonian in Eq.~\eqref{eq:field_hamiltonian} around the mean field solution $\bar{w}$ given in Eq.~\eqref{eq:MFT_w_solution}. To achieve this, we write
\begin{equation*}
w(\rr) = \bar{w} + \phi_0 + \phi(\rr) ,
\end{equation*}
with $\int \dd \rr \phi(\rr) \equiv 0$ and $ \phi_0 $ is the $\kk=\bm{0}$ fluctuation mode. In this field basis, the single molecule partition functions defined in Eqs.~\eqref{eq:polymer_Q} and \eqref{eq:single_bead_Qs} become
\begin{equation*}
\begin{aligned}
Q_{\rm p}[\breve{w},\breve{\psi}] =& \e^{-\ii N(\bar{w} + \phi_0)} Q_{\rm p}[\breve{\phi},\breve{\psi}] , \\
Q_{\rm A,B}[\breve{w},\breve{\psi}] =& \e^{-\ii (\bar{w} + \phi_0)} Q_{\rm A,B}[\breve{\phi},\breve{\psi}] , \\
Q_{\rm AB}[\breve{w},\breve{\psi}] =& \e^{-\ii \gamma (\bar{w} + \phi_0)} Q_{\rm AB}[\breve{\phi},\breve{\psi}] .
\end{aligned}
\end{equation*}
Charge neutrality results in a global shift symmetry $\psi(\rr) \rightarrow \psi(\rr) + ({\rm const.})$ that can be used to eliminate the $\kk=\bm{0}$ mode of $\psi(\rr)$, such that we can assume $\int \dd \rr \psi(\rr) = 0$ in what follows. The single molecule partition functions $Q_i[\breve{\phi},\breve{\psi}]$ have the quadratic expansions
\begin{equation*}
Q_i[\breve{\phi},\breve{\psi}] \approx 1 - \frac{1}{2V} \int \frac{\dd \kk}{(2 \pi)^3} \Psi^{\rm T}(-\kk) \mathsf{G}_i \Psi(\kk) , 
\end{equation*}
for $i={\rm p,A,B,AB} $, with $\Psi(\kk) = \begin{pmatrix}
\tilde{\phi}(\kk) , \tilde{\psi}(\kk)
\end{pmatrix}^{\rm T}$ 
containing the Fourier transformed field fluctuations. Here, $\mathsf{G}_{\rm p}$ is defined as in the main text and
\begin{equation*}
\mathsf{G}_{\rm A,B}  =\hat{\Gamma}^2 \begin{pmatrix}
1 & \pm 1 \\
\pm1 & 1 
\end{pmatrix} , \quad \mathsf{G}_{\rm AB}  =\hat{\Gamma}^2 \begin{pmatrix}
\gamma^2 & 0 \\
0 & 0
\end{pmatrix} . 
\end{equation*}
To find the RPA expansion of the term $\nT h_0(x[w,\psi])$ in the field Hamiltonian, we first express the expansion of the field operator $x[w,\psi]$, defined in Eq.~\eqref{eq:x_def}, as 
\begin{equation*}
x[w,\psi] \approx \bar{x} \e^{(2-\gamma)\ii \phi_0} (1 + \epsilon) \equiv x_0 (1 + \epsilon) , 
\end{equation*}
with $\bar{x} = 2 \rhoT \e^{(2-\gamma) \ii \bar{w}} / K_0$ and $\epsilon \equiv \int \dd \kk \Psi(-\kk)^{\rm T} ( \mathsf{G}_{\rm A} + \mathsf{G}_{\rm B} - \mathsf{G}_{\rm AB} ) \Psi(\kk) / 2 V (2 \pi)^3$. The RPA expansion of $h_0(x)$ simplifies to $h_0(x) \approx h_0(x_0) + h_1(x_0) \epsilon \approx h_0(x_0) + \bar{h}_1 \epsilon$ with $\bar{h}_{i} \equiv h_{i}(\bar{x})$. The expansion of $h_0(x_0)$ to quadratic order in $\phi_0$ becomes $h_0(x_0) \approx \bar{h}_0 + (2-\gamma) \bar{h}_1 \ii \phi_0 - (2-\gamma)^2 \bar{h}_2 \phi_0^2$. This last step can be obtained using the relation $h_0''(x) x^2 = h_2(x)-h_1(x)$ which can be proven by considering $(h_1(x) x)'$ and $h_{i+1}(x) \equiv h'_i(x) x$. 

Combining the above expansion of $h_0(x)$ with the expansion of the other terms in Eq.~\eqref{eq:field_hamiltonian} leads to the RPA expansion of the field Hamiltonian $H[w,\psi] \approx H[\bar{w},0] + H^{({\rm RPA})} $ with
\begin{equation*}
H^{({\rm RPA})} = - \frac{\nT }{2}(2-\gamma)^2 \bar{h}_2 \phi_0^2 + \frac{1}{2} \int \frac{\dd \kk}{(2 \pi)^3} \Psi(-\kk)^{\rm T} \mathsf{G} \Psi(\kk)
\end{equation*}
where $\mathsf{G}$ is defined as in the main text. One can show that the contribution from $\phi_0$ vanishes in the infinite volume limit ($V \rightarrow \infty$), and will therefore be omitted in the following RPA calculations. Note, however, that the quadratic expansion of $H[w,\psi]$ is also used for the semi-implicit CL time integration used in FTS, which accounts for the $\kk=\bm{0}$ modes explicitly.

Performing the Gaussian field integrals over $\phi(\rr)$ and $\psi(\rr)$ leads to the RPA free energy density in Eq.~\eqref{eq:RPA_free_energy}, $f \approx \bar{f} + f^{({\rm fl})}$ with
\begin{equation*}
 f^{({\rm fl})} = \frac{1}{4 \pi^2} \int_0^{\infty} \dd k \, k^2 \ln\left( \frac{4 \pi \lB v}{k^2} \det \mathsf{G} \right) .
\end{equation*}
The derivatives required for the RPA chemical potentials are
\begin{equation*}
\frac{\partial  f^{({\rm fl})} }{\partial \rho_i } = \frac{1}{4 \pi^2} \int_0^{\infty} \dd k \, k^2 {\rm Tr}\left[ \mathsf{G}^{-1} \frac{\partial \mathsf{G} }{\partial \rho_i } \right] , \quad i={\rm b, T}.  \\
\end{equation*}
The density derivatives of the matrix $\mathsf{G}$ includes contributions from the derivatives of the MFT bound fraction $\bar{h}_1$,
\begin{equation*}
\begin{aligned}
\rhoT \frac{\partial \bar{h}_1}{\partial \rhob } &=  \frac{(2-\gamma) \bar{h}_2 \rhoT v}{1-(2-\gamma)^2 \bar{h}_2 \rhoT v } , \\
\rhoT \frac{\partial \bar{h}_1}{\partial \rhoT } &= \bar{h}_2 \left[ 1 + (2-\gamma) \rhoT v \frac{\gamma + (2-\gamma) (\bar{h}_1 + \bar{h}_2) }{1-(2-\gamma)^2 \bar{h}_2 \rhoT v} \right], 
\end{aligned}
\end{equation*}
which can be derived by taking the corresponding derivatives of Eq.~\eqref{eq:MFT_w_solution}. The resulting RPA contributions for the chemical potentials, $\mu_{\rm p}^{({\rm fl})} = N \partial f^{({\rm fl})} / \partial \rhob$ and $\mu_{\rm T}^{({\rm fl})} = \partial f^{({\rm fl})} / \partial \rhoT$, become
\begin{widetext}
\begin{equation*} 
\begin{aligned}
\mu_{\rm p}^{({\rm fl})} &= \frac{N}{4 \pi^2} \int_0^{\infty} \dd k \, k^2 {\rm Tr} \left\lbrace \mathsf{G}^{-1} \left[ \mathsf{G}_{\rm p} +  \frac{(2-\gamma) \bar{h}_2 \rhoT v}{1-(2-\gamma)^2 \bar{h}_2 \rhoT v } (\mathsf{G}_{\rm A} + \mathsf{G}_{\rm B} - \mathsf{G}_{\rm AB} )\right] \right\rbrace , \\
\mu_{\rm T}^{({\rm fl})} &= \frac{1}{4 \pi^2} \int_0^{\infty} \dd k \, k^2 {\rm Tr} \left\lbrace \mathsf{G}^{-1} \left[ \mathsf{G}_{\rm T} + \bar{h}_2 \left( 1 + (2 - \gamma) \rhoT v \frac{\gamma + (2-\gamma) (\bar{h}_1 +  \bar{h}_2) }{1-(2-\gamma)^2 \bar{h}_2 \rhoT v} \right) (\mathsf{G}_{\rm A} + \mathsf{G}_{\rm B} - \mathsf{G}_{\rm AB} ) \right] \right\rbrace .
\end{aligned}
\end{equation*}
\end{widetext}
The fluctuation contribution to the osmotic pressure can then be obtained as $\Pi^{({\rm fl})} = \rhob \mu_{\rm p}^{({\rm fl})} / N + \rhoT  \mu_{\rm T}^{({\rm fl})}  - f^{({\rm fl})}$.

The RPA expression for the density of dissociated ions $[{\rm A}^+]$ follows from Eq.~\eqref{eq:rhofree_from_K0_derivative}, 
\begin{equation*}
[{\rm A}^+] = \bar{h}_1 \rhoT - \frac{1}{4 \pi^2 } \int_0^{\infty} \dd k \, k^2 {\rm Tr} \left[ \mathsf{G}^{-1} K_0 \frac{\partial \mathsf{G} }{\partial K_0} \right], 
\end{equation*}
with
\begin{equation*}
K_0 \frac{\partial \mathsf{G} }{\partial K_0} = - \frac{\rhoT \bar{h}_2  (\mathsf{G}_{\rm A} + \mathsf{G}_{\rm B} - \mathsf{G}_{\rm AB} ) }{1 - (2-\gamma)^2 \rhoT v \bar{h}_2 } . 
\end{equation*}
The resulting RPA formula for the fraction of dissociated ions becomes $\langle h_1(x) \rangle = \bar{h}_1 + h_1^{({\rm fl})}$ with
\begin{equation*}
h_1^{({\rm fl})} = \frac{\bar{h}_2 \int_0^{\infty} \dd k \, k^2 {\rm Tr} \left[ \mathsf{G}^{-1} (\mathsf{G}_{\rm A} + \mathsf{G}_{\rm B} - \mathsf{G}_{\rm AB} ) \right] }{4 \pi^2(1 - (2-\gamma)^2 \rhoT v \bar{h}_2)} , 
\end{equation*}
which can be plugged into Eq.~\eqref{eq:Kd_field_theory} to obtain the RPA dissociation constant.
\section{Detailed expressions of the density operators used FTS}\label{sec:appn_fts}
For a given field configuration $\lbrace w(\rr), \psi(\rr) \rbrace$, the forward and backward propagators $q_{\mathrm{F}}(\rr,\alpha)$ and $q_{\mathrm{B}}(\rr, \alpha)$, respectively,  associated with the discrete bead-spring polymer chain model used in this work can be used to calculate the field operators $\tilde{\rho}_{\mathrm{p}}(\rr)$, $\tilde{c}_{\mathrm{p}}(\rr)$ and $Q_{\mathrm{p}}[\breve{w},\breve{\psi}]$. The chain propagators are constructed iteratively through Kolmogorov--Chapman equations as 
\begin{equation*}
\begin{aligned}
\label{eq:propagators}
q_{\mathrm{F}}(\rr,\alpha+1) &= \e^{-\ii W(\rr,\alpha+1) } \Phi \star q_{\mathrm{F}}(\rr,\alpha)   \, , \\
q_{\mathrm{B}}(\rr,\alpha-1) &= \e^{-\ii W(\rr,\alpha-1)  } \Phi \star q_{\mathrm{B}}(\rr,\alpha) \, ,
\end{aligned}
\end{equation*}
where $W(\rr,\alpha)\equiv \breve{w}(\rr) + \sigma_{\alpha}
\breve{\psi}(\rr)$, $\Phi(\rr) \equiv 
(3 / 2 \pi b^2)^{3/2} \exp(-3 \rr^2 / 2 b^2)$, and
starting from $q_{\mathrm{F}}(\rr,1) = \exp[-\ii W(\rr,1) ]$ and 
$q_{\mathrm{B}}(\rr,N) = \exp[-\ii W(\rr,N)]$.  Here `$\star$' denotes a spatial convolution. One can then show that 
\begin{equation*}
\begin{aligned}
\label{eq:Q_rhos}
Q_{\mathrm{p}}[\breve{w},\breve{\psi}] &= \frac{1}{V} \int \mathrm{d} \rr \, q_{\mathrm{F}}(\rr,N)  \, , \\
\tilde{\rho}_{\mathrm{b}}(\rr) &= \Gamma 
\star \frac{\rho_{\rm b}}{N \, Q_{\mathrm{p}}[\breve{w},\breve{\psi}] } 
\sum_{\alpha=1}^{N}  \e^{\ii W(\rr,\alpha) } q_{\mathrm{F}}(\rr,\alpha) 
q_{\mathrm{B}}(\rr,\alpha) \, , \\
\tilde{c}_{\mathrm{b}}(\rr) &= \Gamma \star \frac{\rho_{\rm b}}{N\, Q_{\mathrm{p}}[\breve{w},\breve{\psi}] } 
\sum_{\alpha=1}^{N}  \e^{\ii W(\rr,\alpha) } q_{\mathrm{F}}(\rr,\alpha) 
q_{\mathrm{B}}(\rr,\alpha) \sigma_{\alpha} \, .
\end{aligned}
\end{equation*}

Bead density operators for $\mathrm{A}^+$, $\mathrm{B}^-$ and $\mathrm{AB}$ are given by
\begin{equation*}
\begin{aligned}
\tilde{\rho}_{\mathrm{A}} &= h_1(x)\frac{\rho_{\mathrm{T}}}{Q_{\mathrm{A}}[\breve{w},\breve{\psi}]} \e^{-\ii \breve{w} - \ii \breve{\psi}}\, ,\\
\tilde{\rho}_{\mathrm{B}} &= h_1(x)\frac{\rho_{\mathrm{T}}}{Q_{\mathrm{B}}[\breve{w},\breve{\psi}]} \e^{-\ii \breve{w} + \ii \breve{\psi}} \, ,\\
\tilde{\rho}_{\mathrm{AB}} &= \left[1 - h_1(x)\right]\frac{\gamma \rho_{\mathrm{T}}}{Q_{\mathrm{AB}}[\breve{w},\breve{\psi}]} \e^{-\ii \gamma \breve{w}} \, . 
\end{aligned}
\end{equation*}

%
\bibliographystyle{unsrt}
\bibliography{refs}

\end{document}